\documentclass[a4paper,11pt]{article}

\usepackage{cite}
\usepackage{epsfig}
\usepackage{amssymb}

\newcommand{\fig}[4]{\begin{figure}[ht]\epsfxsize=#2\bigskip\centerline{\epsfbox{#1}}\caption{\small\it #3 \label{#4}}\bigskip\end{figure}}

\begin{document}

\title{Constraining an Expanding Locally Anisotropic metric from the Pioneer anomaly\\[2mm]}

\author{P. Castelo Ferreira\\[3mm] pedro.castelo.ferreira@gmail.com\\[5mm] \small  Grupo de F\'{\i}sica Matem\'atica -- Universidade de Lisboa\\\small Instituto para a Investiga\c{c}\~ao Interdisciplinar\\\small
Av. Prof. Gama Pinto, 2; 1649-003 Lisboa -- Portugal\\[5mm]\small Eng. Electrot\'ecnica -- FECN\\\small Universidade Lus\'ofona de Humanidades e Tecnologia\\\small Campo Grande 376; 1749-024 Lisboa -- Portugal}

\date{}

\maketitle

\label{firstpage}

\begin{abstract}
It is discussed the possibility of a fine-tuneable contribution to the two way Doppler acceleration either towards, either outwards the Sun for heliocentric distances above $20\,AU$ by considering a background described by an Expanding Locally Anisotropic (ELA) metric. This metric encodes both the standard local Schwarzschild gravitational effects and the cosmological Universe expansion effects allowing simultaneously to fine-tune other gravitational effects at intermediate scales, which may be tentatively interpreted as a covariant parameterization of either cold dark matter either gravitational interaction corrections. Are derived bounds for the ELA metric functional parameter by considering the bounds on the deviation from standard General Relativity imposed by the current updated limits for the Pioneer anomaly, taking in consideration both the natural outgassing and on-board radiation pressure, resulting in an average Doppler acceleration outwards the Sun of $a_p\approx +0.4^{+2.1}_{-2.0}\times 10^{-10}\,(m\,s^{-2})$. It is also computed the mass-energy density for the ELA metric within the bounds obtained and are discussed the respective contributions to the cosmological mass-energy density which, for compatibility with the $\Lambda CDM$ model, are included in $\Omega_{CDM}$.
\end{abstract}

\ \vfill\ \\

keywords: gravitation, Pioneer, interplanetary medium, equation of state, dark matter, cosmological parameters

\thispagestyle{empty}
\newpage
\setcounter{page}{1}
\section{Introduction}

The Pioneer space-crafts have been launch from Earth on 1972 and 1973 travelling outwards of the Solar system~\cite{Pioneer_1a,Pioneer_1b,Pioneer_2}.
In between the heliocentric distances of $20\,AU$ and $70\,AU$, travelling at an approximately constant velocity of $v_p=12.2\times 10^3\,m\,s^{-1}$, it was detected an unmodeled Doppler shift known as the Pioneer effect. Originally attributed to a physical acceleration, this effect was interpreted as a constant radial acceleration towards the Sun of $a_p=-(8.74\pm 1.33)\times 10^{-10}\,m\,s^{-2}$~\cite{Pioneer_2}.
Many possible explanations for this effect have been considered in the literature,
including modified gravity theories and models~\cite{mod_grav_p1}, non-gravitational interactions~\cite{extra_matter1}, extra-dimensional theories
and models~\cite{extra_dimensions1} and dark matter~\cite{dm_p1}.

However recently the Pioneer data was re-analysed favouring a significant jerk in the measured Doppler acceleration such that the best exponential fit is approximately given as a function of the radial distance to the Sun (for $r \in ]20AU,70AU[$) by~\cite{toth_1,toth_2}
\begin{equation}
a_{exp}(r)\approx-12.22^{+0.16}_{-0.16}\times 10^{-10}e^{-\frac{2.599\times 10^{-12}\,\log(2)}{28.8\pm 0.07}\,\left(r-1.496\times 10^{11}\right)}\ (m\,s^{-2})\ .
\label{ap}
\end{equation}
In addition it has been put forward that this acceleration is mainly due to two distinct effects.
The more significant contribution $a_T$ is of thermal origin, the antenna located at the back of the space-craft works as a radiation sail propelled by the radiation emissions from the on-board power source and the space-craft electronics~\cite{rad_press_1,rad_press_2,rad_press_3}. Also the natural outgassing of the space-craft material~\cite{rad_press_3} contributes a significant amount $a_L$ for the Pioneer acceleration. These two contributions are (for $r \in ]20AU,70AU[$)~\cite{rad_press_3}  
\begin{equation}
\begin{array}{rcl}
a_T(r)&\approx &-10.7^{+2.4}_{-2.4}\times 10^{-10}e^{-6.238\times 10^{-14}\left(r-1.496\times 10^{11}\right)}\ (m\,s^{-2})\ ,\\[3mm]
a_L(r)&\approx &-7.15^{+3.15}_{-2.50}\times 10^{-11}\,\left(1+\frac{2.389\times 10^6}{\sqrt{r}}\right)\ (m\,s^{-2})\ .
\end{array}
\label{ap_L_T}
\end{equation}
The resulting acceleration in the range $r \in ]20AU,70AU[$ is, within the experimental plus the estimative error bars, null. Its average value in this range is
\begin{equation} 
\left<a_p(r)\right>=\left<a_{exp}(r)-a_L(r)-a_T(r)\right>\approx 0.4^{+2.1}_{-2.0}\times 10^{-10}\,(m\,s^{-2})\ .
\label{a_p}
\end{equation}
The several contributions for the unmodeled acceleration of Pioneer 10 and the resulting
acceleration are plotted in figure~\ref{fig.ap}.
\fig{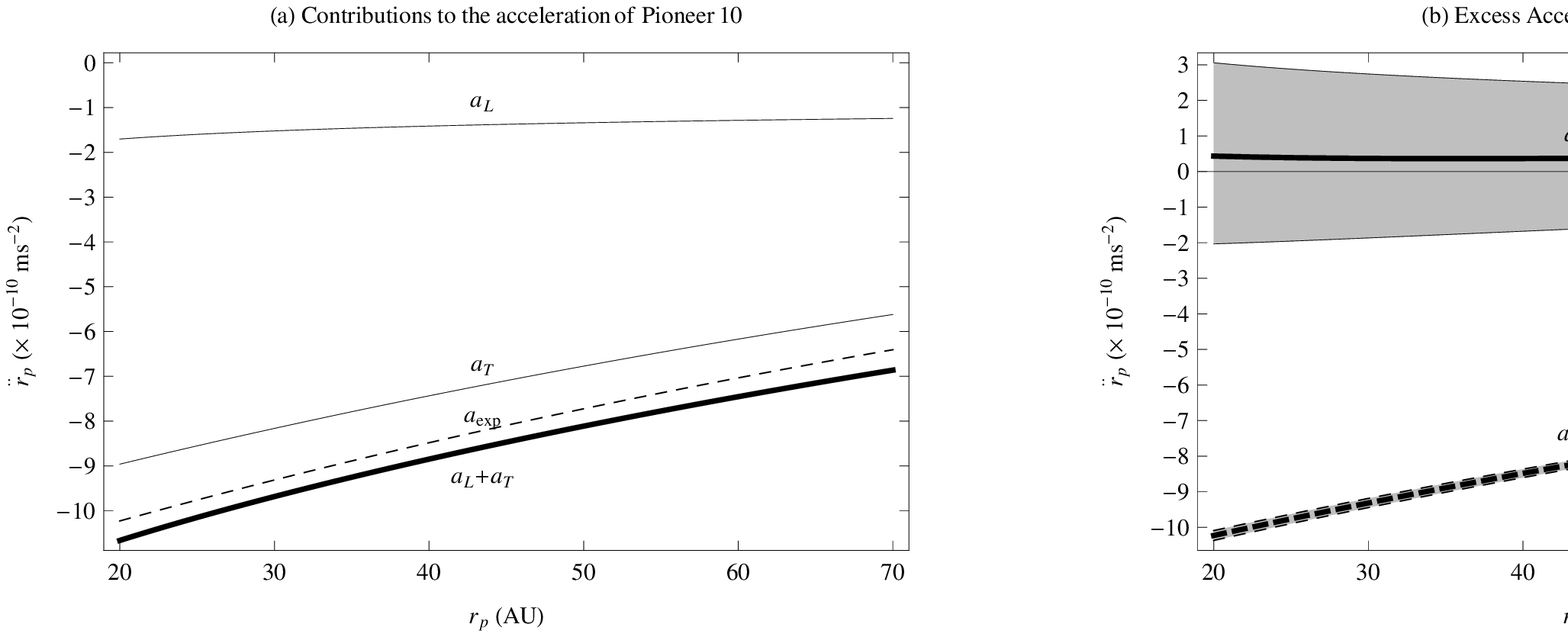}{90mm}{{\bf (a)} contributions $a_L(r)$ and $a_T(r)$~(\ref{ap_L_T}) to the unmodeled acceleration $a_{exp}(r)$~(\ref{ap}) of Pioneer 10; {\bf (b)} total Pioneer 10 acceleration $a_p(r)=a_{exp}(r)-a_L(r)-a_T(r)$. The shaded region represents the quoted experimental plus
estimative error bars~\cite{toth_1,toth_2,rad_press_2,rad_press_3}.}{fig.ap}

The main purpose of the this paper is to investigate whether a contribution to the Doppler acceleration of the Pioneer space-craft can be consistently obtained considering a background described by the ELA metric~\cite{PLB,e-print}, an anisotropic~\cite{anisotropy} ansatz that generalizes the isotropic McVittie metric~\cite{McV1,McV2,McV3,McV4} interpolating between the Schwarzschild (SC) metric~\cite{Schwarzschild} near massive bodies and the Robertson-Walker (RW) metric~\cite{RW1} at spatial infinity, hence describing local matter in an expanding Universe~\cite{Hubble1}. In particular if the bounds on the Pioneer 10 acceleration discussed above can be employed to constraint the parameters of such metric. 

The shift function of the ELA metric can be fine-tuned through an exponential functional parameter allowing
for a covariant parameterization of intermediate scale gravitational effects~\cite{e-print,curves}, hence
allowing to describe either local cold dark matter effects~\cite{DM1,DM2} either modified theories of gravity~\cite{mod_grav1}. Although this fine-tuning is technically more challenging than semi-classical heuristic approaches it has the following advantages:
\begin{itemize}
\item it is a covariant formulation allowing to deal with gravitational corrections within the framework of General Relativity;
\item in particular allows for direct covariant computations of gravitational accelerations and gravitational frequency-shifts;
\item allows as well for QFT cross-section computations in curved backgrounds to include cold dark matter effects;
\item it allows for the description of intermediate scale gravitational effects maintaining the local and cosmological physical properties of space-time.
\end{itemize}

In the following we will employ two distinct coordinate systems, expanding coordinates, $r_1$, for which the area of the sphere is $A=4\pi\,r_1^2$
such that the coordinate quantities correspond directly to
measurable lengths and the standard Robertson-Walker coordinates, $r$, for which the area of the sphere is $A=4\pi\,a^3r^2$ being suitable for direct computations of measurable red-shifts~\cite{e-print}. The map between both these coordinate systems is $r_1=a\,r$ and the ansatz for the ELA metric is~\cite{PLB,e-print} 
\begin{equation}
\begin{array}{rcl}
ds^2&=&\displaystyle(1-U)(cdt)^2-d\Omega_1^2\\[3mm]
&&-\left(\frac{dr_1}{\sqrt{1-U}}-\frac{H\,r_1}{c}(1-U)^{\frac{\alpha}{2}}(cdt)\right)^2\\[5mm]
&=&\displaystyle(1-U)(cdt)^2-a^2d\Omega^2\\[3mm]
&&-a^2\left(\frac{dr}{\sqrt{1-U}}+\frac{H\,r}{c}\left(1-(1-U)^{\frac{\alpha}{2}}\right)(cdt)\right)^2\ ,
\end{array}
\label{ELA_metric}
\end{equation}
where $a$ is the universe scale factor, $H=\dot{a}/a$ is the time dependent Hubble rate, $U=2GM/(c^2\,r_1)=2GM/(c^2\,ar)$ is the standard SC gravitational potential, $M$ is the standard SC mass (the value of the gravitational mass pole), 
$c$ is the speed of light, $G$ is the Gravitational constant and the solid angle line-elements are $d\Omega_1^2=r_1^2(d\theta^2+\sin^2\theta d\varphi^2)$ and $d\Omega^2=r^2(d\theta^2+\sin^2\theta d\varphi^2)$.

This ansatz generalizes previous ansatz and solutions in the literature~\cite{McV1,McV2,McV3,McV4} allowing for local anisotropy~\cite{anisotropy} and having the novelty of describing a
space-time without singularities at the SC radius simultaneously maintaining
as asymptotic limits the SC metric near the SC event horizon and
the RW metric at spatial infinity. To ensure these properties of space-time the following limits must be obeyed~\cite{PLB,e-print}
\begin{equation}
\begin{array}{ll}
\displaystyle\lim_{r_1\to 0}(1-U)^\alpha=0&\mathrm{:\ SC\ mass\ pole\ \mathit{M}\ is\ maintained}\ ,\\[3mm]
\displaystyle\lim_{r_1\to r_{1.SC}}\alpha\ge 3&\mathrm{:\ space-time\ is\ singularity\ free}\\[2mm]
&\hfill\mathrm{at\ SC\ horizon}\ ,\\[3mm]
\displaystyle\lim_{r_1\to \infty}(1-U)^\alpha= 1&\mathrm{:\ metric\ converges\ to\ RW\ metric}\ .
\end{array}
\label{alpha_lim}
\end{equation}

The main drawback of this metric is the absence of a direct physical interpretation for the functional parameter $\alpha$. Here we will not attempt to solve this problem, we recall that the ELA metric was originally introduced phenomenologicaly in~\cite{PLB,e-print} to generalize the McVittie metric~\cite{McV1} and it is at most interpreted as a heuristic fine-tuneable parametrization of gravitational interactions at intermediate scales. Hence we consider a branch ansatz with the following properties:
\begin{itemize}
\item allows the limits~(\ref{alpha_lim}) to be obeyed both near the origin and at spatial infinity;
\item has no relevant contributions for planetary motion within the solar system;
\item allows to fit the Doppler acceleration for the heliocentric distance of $20\,AU$ up to $70\,AU$;
\item allows to model the mass-energy density described by this background for distances above $70\,AU$ up to some upper cut-off distance $R_{max}$.
\end{itemize}
In particular, above the cut-off $R_{max}$, we explicitly consider the McVittie metric, hence $\alpha=0$, such that explicitly isotropy is retrieved for large radial distances and the mass-energy density for the background metric coincides with the one for the RW metric. Also we do not discuss here possible effects on
planetary motion, leaving an account of this problem to a future work. Here for heliocentric distances below $20\,AU$ we consider the lower possible value of $\alpha\approx 3$ compatible with the limits~(\ref{alpha_lim}) such that the gravitational effects due to the background are negligible for planetary dynamics within the inner Solar system.
Therefore we consider the following expression for the functional parameter $\alpha$:
\begin{equation}
\alpha(r_1)=\left\{
\begin{array}{lcl}
(3-\alpha_1)+\alpha_1\,U&,&0\leq r_1< 20\,AU\\
\displaystyle\alpha_2(r)&,&20\,AU\leq r_1\leq R_{max}\\
0&,&r_1>R_{max}
\end{array}
\right.
\label{alpha}
\end{equation}
where $-1\ll\alpha_1<0$ is a negative constant arbitrarily close to zero that ensures that the SC mass
pole is maintained at the origin~\cite{PLB} being its effects negligible far from the SC horizon, $\alpha(2GM/c^2\ll r_1<20\,AU)\approx 3$.
As for $\alpha_2(r)$ is a generic function that can be fitted to the Pioneer
experimental data above the heliocentric distance of $20\,AU$.

We have organized this work as follows. In section~\ref{sec.Doppler} are derived the Doppler
shift for a two way range Doppler measurement as well as the gravitational acceleration
correction for the background described by the ELA metric~(\ref{ELA_metric}). Then in section~\ref{sec.rho} are derived the mass-energy density and the pressures as well as the mass corrections due to the
extended configurations corresponding to the functional parameter ansatz~(\ref{alpha}). It is also
discussed the respective contributions to the relative cosmological mass-energy densities within the
$\Lambda CDM$ model~\cite{wmap}. In section~\ref{sec.bounds} are computed and discussed several fits to the functional parameter $\alpha$ within the error bars of the Pioneer 10 data. Finally in the conclusions are resumed the results obtained and discussed its relevance as well as future research topics.

\section{Doppler shift and gravitational acceleration\label{sec.Doppler}}

The Pioneer effect was measured using a two-way Doppler shift~\cite{Pioneer_1a,Pioneer_1b}. The standard gravitational red-shift as described by the SC metric is null, as for the standard cosmological red-shift is negligible as noted in~\cite{Pioneer_2,e-print} being below the accuracy of the Pioneer measurements.
For this reason the Pioneer effect is often attributed to a correction to the gravitational acceleration~\cite{mod_grav_p1,extra_dimensions1,dm_p1}.

Taking as central mass the Sun and neglecting the contribution of other bodies in the solar system, the corrections to the gravitational acceleration felt by a test mass travelling in the background given by the ELA metric~(\ref{ELA_metric}) with respect to the General Relativity gravitational
acceleration corresponding to a Ricci-flat background described by the SC metric 
is~\cite{e-print,curves}
\begin{equation}
\begin{array}{rcl}
\delta\ddot{r}_p&\approx&\displaystyle\frac{H^2r_p}{2}\,\left(1-U_{\odot}\right)^{\alpha}\left(2(1-U_{\odot})-(1+\alpha)r_p\,U_{\odot}'\right.\\[5mm]
&&\displaystyle-2(1+q)(1-U_{\odot})^{\frac{1}{2}-\frac{\alpha}{2}}\\[5mm]
&&\displaystyle\left.+r_p(1-U_{\odot})\ln(1-U_{\odot})\alpha'\right)+O(H^4)\ ,
\end{array}
\label{F}
\end{equation}    
where the primed quantities stand for differentiation with respect to the radial coordinate $r_1$, $r_p$ stands for the geometrical (measurable) distance from the Sun to the Pioneer space-craft and $U_{\odot}=2GM_{\odot}/(c^2r_p)$ stands for the standard SC gravitational potential of the Sun with
$M_{\odot}=1.9891\times 10^{30}\,kg$.

In addition there is a correction to the gravitational red-shift. The measurable
red-shift due to the background is obtained from the line element~(\ref{ELA_metric}). For a light ray travelling approximately at a radial trajectory with respect to the Sun described by the four-vector $k^\mu\approx(\omega/c,k^r,0,0)$ , from earth at $r^{(0)}_1=a_0r^{(0)}\approx 1\,AU$ to the Pioneer space-craft at $r^{(1)}_1=a_1\,r^{(1)}=r_p$ and back to earth at $r^{(2)}_1=a_2\,r^{(2)}\approx 1\,AU$ we obtain that the ratios
between the radial wave numbers of the emitted radiation from earth
$k^r_{(0)}$, the radiation received at the Pioneer space-craft $k^r_{(1)}$
and the radiation received back at Earth $k^r_{(2)}$ are
\begin{equation}
\begin{array}{rcl}
\displaystyle\frac{k^r_{(1)}}{k^r_{(0)}}&=&\displaystyle\frac{a_0}{a_1}\,\frac{1-U_p-\delta}{1-U_e}\ ,\\[5mm]
\displaystyle \frac{k^r_{(2)}}{k^r_{(1)}}&=&\displaystyle\frac{a_1}{a_2}\,\frac{1-U_e}{1-U_p+\delta}\ ,\\[5mm]
\displaystyle \delta&=&\displaystyle \frac{H\,r_p}{c}\left(1-(1-U)^{\frac{\alpha}{2}+\frac{1}{2}}\right)\ ,
\end{array}
\end{equation}
where $U_p$ and $U_e$ are the gravitational potential of the Sun evaluated
at the Pioneer space-craft and at Earth, and $a_0$, $a_1$ and $a_2$ stand for
the Universe scale factor $a=a(t)$ evaluated at the time of emission of the radiation from Earth, the time of receiving and re-emission at the Pioneer space-craft and the time of receiving it back at Earth, respectively.
Further noting that expanding the
Universe scale factor to first order in the Hubble rate we obtain that
$a_1=a_0(1+(H/c)r_p)+O(H^2)$ and $a_2=a_0(1+2(H/c)r_p)+O(H^2)$, the measurable
frequency-shift (red-shift if positive or blue-shift if negative) due to the background for a two
way Doppler measurement from Earth to Pioneer and back is
\begin{equation}
\begin{array}{rcl}
\displaystyle\frac{\Delta\nu}{\nu_0}&=&\displaystyle\frac{k^r_{(1)}}{k^r_{(0)}}\,\frac{k^r_{(2)}}{k^r_{(1)}}-1\\[5mm]
&=&\displaystyle\frac{1}{1+2\frac{H\,r_p}{c}}\,\frac{1-U_p-\delta}{1-U_p+\delta}-1+O(H^4)
\end{array}
\end{equation}
Hence the contribution to the Pioneer non-physical acceleration due
to the background red-shift is obtained by differentiating the standard
two way Doppler shift $\Delta\nu/\nu_0=-2\dot{r}_p/c$
\begin{equation}
\delta\ddot{r}_p^{Doppler}=-c\,\frac{v_p}{2}\,\left(\frac{\Delta\nu}{\nu_0}\right)'\ ,
\label{F_Doppler}
\end{equation}
where the prime denotes differentiation with respect to the radial geometric distance (the coordinate $r_1$), $v_p=\dot{r}_p$ for the radial geometric
velocity of the space-craft (obtained from the differentiation $\partial_{t}=\partial_tr_1\,\partial_{r_1}$) and we have neglected relativistic
corrections ($\gamma\approx 1$ for $v_p\ll c$).

Before proceeding to fit these results to the Pioneer acceleration bounds represented in figure~\ref{fig.ap}  we will further derive and discuss the mass-energy density of the background described by the ELA metric~(\ref{ELA_metric}).

\section{Mass-energy density and pressures\label{sec.rho}}

The mass-energy density, $\rho_{(\alpha)}$ and anisotropic
pressures $p_{r(\alpha)}$ and $p_{\theta(\alpha)}=p_{\varphi(\alpha)}$ for the ELA metric~(\ref{ELA_metric}) are
\begin{equation}
\begin{array}{rcl}
\rho_{(\alpha)}&=&\displaystyle\frac{H^2}{8\pi\,G}\left(1-U\right)^{\alpha-1}\left(\alpha\,U+(1-U)\times\right.\\[2mm]
&&\hfill\displaystyle\times\left.\left(3+r_1\log\left(1-U\right)\partial_{r_1}\alpha\right)\right)\ ,\\[5mm]
p_{r(\alpha)}&=&\displaystyle\frac{c^2\,H^2}{8\pi\,G}\left(1-U\right)^{\frac{\alpha-1}{2}}\left(2(1+q)-\left(1-U\right)^{\frac{\alpha-1}{2}}\left(\alpha\,U\right.\right.\\[2mm]
&&\displaystyle\hfill\left.+(1-U)(3+r_1\log\left(1-U\right)\partial_{r_1}\alpha\right)\bigg)\ ,\\[5mm]
p_{\theta(\alpha)}&=&\displaystyle\frac{c^2\,H^2}{16\pi\,G}\left(1-U\right)^{\frac{\alpha-3}{2}}\Bigg[4(1+q)\left(1-U\left(1-\frac{\alpha}{4}\right)\right)\\[2mm]
&&\displaystyle+2\left(1-U\right)^{\frac{\alpha-1}{2}}\Big(3U\left(2-U\left(1-\frac{\alpha}{2}\right)\right)\left(1-\frac{\alpha}{3}\right)\\[2mm]
&&\displaystyle-3-r_1U(1-U)\partial_{r_1}\alpha\Big)+r_1(1-U)\log(1-U)\times\\[2mm]
&&\displaystyle\times\Big((1+q)\partial_{r_1}\alpha-2U\left(1-U\right)^{\frac{\alpha-1}{2}}\alpha\partial_{r_1}\alpha\\[2mm]
&&\displaystyle-\left(1-U\right)^{\frac{\alpha+1}{2}}\Big(6\partial_{r_1}\alpha+r_1\log(1-U)(\partial_{r_1}\alpha)^2\\[2mm]
&&\displaystyle\hfill+r_1\partial^2_{r_1}\alpha\Big)\Big)\Bigg].
\end{array}
\label{rho}
\end{equation}
When the limits~(\ref{alpha_lim}) are obeyed~\cite{PLB}, at $r_1=r_{1.\mathrm{SC}}$, $\rho_{(\alpha)}$ coincides with
the mass-energy density for Ricci flat space-times, being null $\rho_{(\alpha)}(r_1\to r_{1.\mathrm{SC}})=0$, and at spatial infinity it asymptotically converges to
the RW mass-energy density $\rho_{(\alpha)}(r_1\to +\infty)=\rho_{RW}=3\,h^2/(8\pi\,G)$, as well as $p_{r(\alpha)}(r_1\to +\infty)=p_{\theta(\alpha)}(r_1\to +\infty)=p_{\varphi(\alpha)}(r_1\to +\infty)=p_{RW}=(2q-1)\,h^2/(8\pi\,G)$ such that at spatial infinity the RW equation of state is obtained $\omega_{RW}=(2q-1)/3= -0.728$ (this value is obtained within the $\Lambda$CDM model assuming that the pressure of the background is given only by the cosmological constant~\cite{wmap}). Hence this mass-energy distribution has been interpreted in~\cite{PLB}, for a constant functional exponent $\alpha$ as a local anisotropic extended correction to the global
isotropic cosmological background due to a local point mass vanishing at spatial infinity.

In addition to preserve causality it is required that the mass energy density be positive, hence $\rho_{(\alpha)}\geq 0$. This inequality is bounded by the solution of the differential equation $\alpha(r_1)=-(1-U(r_1))(3+r_1\log(1-U(r_1))\alpha'(r_1))/U(r_1)$ holding the following inequality
\begin{equation}
\alpha(r_1)\geq\frac{\bar{\alpha}_{const}-3\log(r_1)}{\log(1-U(r_1))}\ ,
\label{max_bound}
\end{equation}
where $\bar{\alpha}_{const}$ is an arbitrary integration constant.

With respect to the contribution to the cosmological mass-energy density due to the ELA metric, it can be estimated by comparing the total mass contribution of the mass-energy density $\rho_{(\alpha)}$~(\ref{rho}) with the central mass contribution $M_{\odot}$. To compute such contribution it is enough to integrate the difference between the $\rho_{(\alpha)}$ and
the RW expanding background mass-energy density $\rho_{RW}=3\times H^2/(8\pi G)$. Specifically for the branch ansatz~(\ref{alpha}) we obtain
\begin{equation}
\begin{array}{rcl}
\displaystyle M_{\alpha}&=&\displaystyle 4\pi\int_{0}^{+\infty}r_1^2(\rho_{(\alpha)}-\rho_{RW})dr_1\\[5mm]
&=&\displaystyle \frac{H^2}{2\,G}\left[\left(\left(1-U(R_{max})\right)^{\alpha(R_{max})}-1\right)R_{max}^3\right.\\[5mm]
&&\displaystyle+\left(\left(1-U(20\,AU)\right)^{\alpha(20^-\,AU)}-1\right)(20\,AU)^3\\[5mm]
&&\displaystyle\left.-\left(\left(1-U(20\,AU)\right)^{\alpha(20^+\,AU)}-1\right)(20\,AU)^3\right] ,
\end{array}
\label{array}
\end{equation}
where the last two lines are the contributions due to the discontinuity at $r_1=20\,AU$ and above $R_{max}$ the ansatz exactly matches the McVittie metric ($\alpha=0$) such that $\rho_{(\alpha)}=\rho_{RW}$ and the mass contribution is null. We note that as long as the limits~(\ref{alpha_lim}) are obeyed near the origin,
besides the standard SC mass pole $M_{\odot}$, there is no further mass contribution due to the divergence at
the origin. In addition convergence of $M_{\alpha}$ when the cutoff $R_{max}$ is taken to infinity ($R_{max}\to+\infty$) requires more severe conditions than the limits~(\ref{alpha_lim}), in particular it can be fine-tuned either to be null (e.g. for $\alpha_2(r_1\to +\infty)={\mathrm{constant}}/r_1^n,\,n>2$) or to a multiple of the central mass (e.g. for $\alpha_2(r_1\to +\infty)={\mathrm{constant}}/r_1^2$ we obtain $M_{\alpha}=H^2\,{\mathrm{constant}}\,M_\odot/(2G)$). Nevertheless keeping $R_{max}$ finite in the ansatz~(\ref{alpha}) keeps the quantity $M_{\alpha}$
finite and allows to fine-tune the total mass of the background.

For exemplification purposes the contribution to the cosmological mass-energy density due to $M_{\alpha}$ can be estimated by assuming that all barionic matter has a proportional contribution to the one obtained here for
the Sun such that $\Omega_{\alpha}/\Omega_{b}=M_{\alpha}/M_{\odot}$, hence
\begin{equation}
\Omega_{\alpha}=\frac{M_{\alpha}}{M_{\odot}}\,\Omega_{b}\ ,
\label{cosm_den}
\end{equation} 
where $\Omega_b=0.0456$ is the relative cosmological barionic mass-energy within the $\Lambda CDM$ model~\cite{wmap}. The contributions to $\Omega_b$ are given mostly by detectable massive bodies in the universe, hence a direct estimation is possible. Also we recall that within the $\Lambda CDM$ model the remaining contributions to the total mass-energy density are due to the cosmological constant $\Omega_\Lambda=-(2q-1)/3=0.728$ and due to non detectable cold dark matter $\Omega_{CDM}=0.226$ (also denominated $\Omega_c$). The radiation density as well as the curvature density are negligible such that the total relative density is $\Omega=\Omega_b+\Omega_{CDM}+\Omega_\Lambda=1$. With respect to the cosmological equation of state we note that both the barionic matter and cold dark matter have null pressures, while the cosmological constant has a negative pressure with equation of state $\omega_\Lambda=-1$. Then the only contribution to the cosmological pressure is due to the cosmological constant such that the equation of state of the Universe is  $\omega=p_\Lambda/\Omega=-\Omega_\Lambda=(2q-1)/3=-0.728$. Further noting that, due to the upper cut-off $R_{max}$ in the ansatz for $\alpha$~(\ref{alpha}), the average contributions to the cosmological pressures due to the ELA metric background are null
\begin{equation}
\begin{array}{rcl}
\displaystyle\left<p_r\right>_{r\in]0,\infty[}&=&\displaystyle\lim_{\Delta\,r\to +\infty}\frac{\displaystyle 4\pi\int^{R_{max}}_0 r^2\,p_r\,dr}{\Delta\,r}=0\ ,\\[5mm]
\displaystyle\left<p_\theta\right>_{r\in]0,\infty[}&=&\displaystyle\lim_{\Delta\,r\to +\infty}\frac{\displaystyle 4\pi\int^{R_{max}}_0 r^2\,p_\theta\,dr}{\Delta\,r}=0\ ,
\end{array}
\label{ave_pressures}
\end{equation}
we conclude that, within the $\Lambda CDM$ model, contributions to the cosmological densities from the background~(\ref{ELA_metric}) with the functional parameter anzatz given in~(\ref{alpha}) must me included
in $\Omega_{CDM}$.

Possible alternatives to the ansatz~(\ref{alpha}) that consider non-null values ($\alpha\neq 0$) for the functional parameter up to spatial infinity would also allow for non-null contributions to the cosmological
pressure hence also allowing for a contribution to both the cosmological constant density and pressure. We do not develop this
construction here.

As for the local equations of state for the ELA metric we note that generally the pressures $p_{r(\alpha)}$ and $p_{\theta(\alpha)}=p_{\varphi(\alpha)}$ are negative far from the SC event horizon, in particular for constant values of the parameter $\alpha\geq 3$, there are regions where both $\omega_{r(\alpha)}=p_{r(\alpha)}/(c^2\rho_{(\alpha)})$ and $\omega_{\theta(\alpha)}=p_{\theta(\alpha)}/(c^2\rho_{(\alpha)})$ are above $-1$ and regions where $\omega_{r(\alpha)}$ is above $-1$ and $\omega_{\theta(\alpha)}=p_{\theta(\alpha)}/(c^2\rho_{(\alpha)})$ is 
below $-1$~\cite{PLB}. The interpretation of the matter content for these two regions is distinct,
 in the first regions the matter distribution may be interpreted as due
to a local configuration of a standard scalar field ($\omega_{r(\alpha)}>-1$ and $\omega_{\theta(\alpha)}=\omega_{\varphi(\alpha)}>-1$)~\cite{scalar1}, while
in the second regions the matter distribution is compatible with a standard scalar field along the radial coordinate direction ($\omega_r>-1$), along the angular variables directions it is obtained an equation of state commonly attributed to Phantom matter ($\omega<-1$)~\cite{Phantom_}. We recall that the main difference between these two scalar fields is the sign of the kinetic term in the action, hence a implementation for these regions requires that the kinetic term along the angular variables directions to be negative while being positive along the radial coordinate directions. Possibly the simpler implementation for these equations of state is to consider a local configuration for a radially symmetric anisotropic scalar field. Such model may be compatible, for instance, with locally anisotropic string theories and string inspired cosmological models~\cite{string}.  

Next we will employ the Pioneer bounds on $\ddot{r}_p$ to obtain bounds on the parameter $\alpha$
above the heliocentric distance of $20\,AU$ and discuss the total contributions of such configurations to the cosmological mass-energy densities.

\section{Bounds on $\alpha$\label{sec.bounds}}

A contribution to the Pioneer Doppler acceleration due to the background encoded in the metric~(\ref{ELA_metric}) is the sum of the contribution~(\ref{F}) and (\ref{F_Doppler})
\begin{equation}
a_{\alpha}=\delta\ddot{r}_p+\delta\ddot{r}_p^{Doppler}\ .
\end{equation}
Next we consider and analyse 4 distinct fits of the functional parameter $\alpha_2(r_p)$ within the acceleration bounds for $a_p$ as represented by the shaded region in figure~\ref{fig.ap}:
\begin{itemize}
\item[$\alpha_{max}$] the allowed maximum valued fit for $\alpha$ corresponding to positive Doppler acceleration $a_{\alpha_{max}}>0$ as close as possible from the upper boundary of the allowed region;
\item[$\alpha_+$] a fit for $\alpha$ corresponding to positive Doppler acceleration $a_{\alpha_{+}}>0$ half way the positive acceleration allowed region;
\item[$\alpha_-$] a fit for $\alpha$ corresponding to negative Doppler acceleration $a_{\alpha_{-}}<0$ half way the negative acceleration allowed region;
\item[$\alpha_{min}$] the allowed minimum valued fit for $\alpha$ corresponding to negative Doppler acceleration $a_{\alpha_{min}}<0$ as close as possible from the from the lower boundary of the allowed region. 
\end{itemize}
First let us note that for positive values of the function $\alpha_2(r_p)$ the effects are of the same order
of magnitude of the pure background expansion, $a_\alpha\sim+10^{-14}\,m\,s^{-2}$, hence to obtain
measurable effects it is required to consider negative values for this parameter.

We remark that an exact fit to the upper boundary of the allowed positive Doppler acceleration $a_\alpha$ correspond to strictly negative values for the mass-energy density $\rho_{(\alpha)}$ in the range $r_p\in]20\,AU,70\,AU[$, therefore for the fit $\alpha_{max}<0$ we consider the limiting solution~(\ref{max_bound}) with $a_{\alpha_{max}}(20\,AU)=3.057\times 10^{-10}\,m\,s^{-2}$, hence with $\bar{\alpha}_{const}=106.176$. As for the fit $a_{\alpha_{+}}$, for exemplification purposes and to ensure positive $\rho_{(\alpha)}$ we consider the fitting function to be $9/10$ of the limiting solution~(\ref{max_bound}) with $a_{\alpha_{max}}(20\,AU)=1.529\times 10^{-10}\,m\,s^{-2}$, hence with $\bar{\alpha}_{const}=107.650$.
As for the $\alpha_{-}$ and $\alpha_{min}$ corresponding to negative values for the Doppler acceleration $a_{\alpha}<0$ we consider a power series fit given by the following expression
\begin{equation}
\alpha_2(r_p)=\sum^{N}_{n=-1}\alpha_{2.n}\,r_p^n\ .
\label{polinomial_fit}
\end{equation}
For the specific case of the $\alpha_{-}$ and $\alpha_{min}$ we consider $N=10$ and the specific coefficients are listed in appendix~\ref{sec.app}. The 4 fits considered are pictured in figure~\ref{fig.fits}~(a). Noting
that $\alpha_{min}$ and $\alpha_{max}$ define the lower (negative) and upper (positive) limits on the Doppler acceleration allowed by the bounds on the Pioneer 10 acceleration represented in figure~\ref{fig.ap} we conclude that in the range $r_p\in]20\,AU,70\,AU[$ the functional parameter $\alpha_2$ has the lower bound represented in figure~\ref{fig.fits}~(b).
\fig{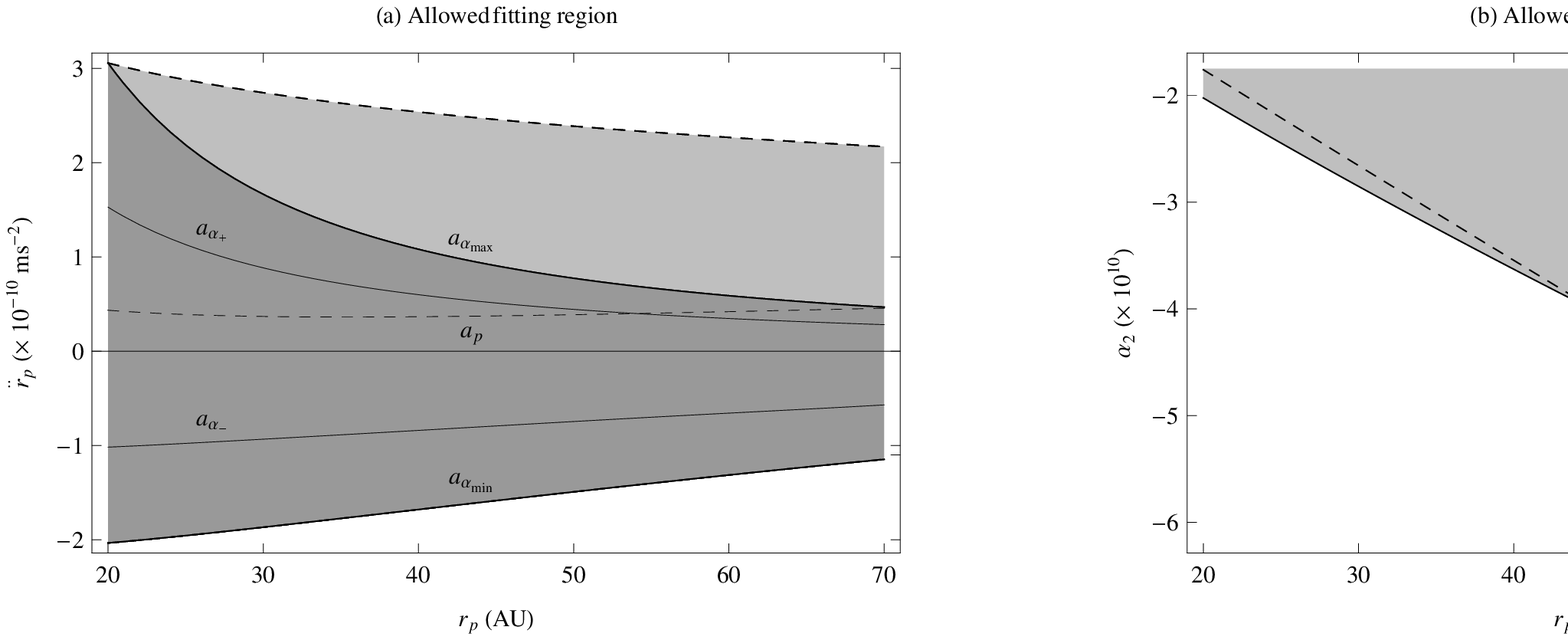}{90mm}{Fit of the functional parameter $\alpha_2(r_1)$~(\ref{alpha}) to the Pioneer acceleration bounds of figure~\ref{fig.ap}: (a) $\alpha_{max}$ corresponds to the limiting solution~(\ref{max_bound}) with $\bar{\alpha}_{const}=106.176$, the solution $\alpha_{+}$ to $9/10$'s of the limiting solution~(\ref{max_bound}) with $\bar{\alpha}_{const}=107.650$, the solution $\alpha_{-}$ and $\alpha_{min}$ correspond to a power series fit of the parameter $\alpha$~(\ref{polinomial_fit}); (b) lower bounds for $\alpha<0$.}{fig.fits}

The dominant contribution to $a_\alpha$ for the above fits between heliocentric distances of $r_p\in]20\,AU,70\,AU[$ is the background frequency-shift, $\delta\ddot{r}_p^{Doppler}\sim -10^{-10}\,ms^{-2}$. As for the value of the physical gravitational acceleration, it is 4 to 5 orders of magnitude below this value, for $\alpha_{max}$ and $\alpha_{+}$ we obtain $\delta\ddot{r}_p\sim -10^{-14}\,ms^{-2}$ (towards the Sun) and for $\alpha_{min}$ and $\alpha_{-}$ we obtain $\delta\ddot{r}_p\sim +10^{-15}\,ms^{-2}$ (outwards the Sun). These values are well below the standard
General Relativity acceleration which is of order $\ddot{r}_{GR}\sim -10^{-6}\,ms^{-2}$.
Nevertheless the effects of this correction to the gravitational acceleration can be measurable at least for Pluto and Neptune~\cite{planets1}, for instance if this correction is fully accounted by an orbital radius variation
we would obtain at an orbital radius of $r_{orb}=20\,AU$ for $\alpha_{max}$ and $\alpha_+$ the value $\dot{r}_{orb}/r_{orb}\sim -10^{-28}\,s^{-1}$ and for $\alpha_{min}$ and $\alpha_-$  the value $\dot{r}_1/r_1\sim +10^{-29}\,s^{-1}$. These values are well below the experimentally measured orbital radius variation for inner planets in the Solar system such as Venus and Mars, $|\dot{r}_{orb}/r_{orb}|< 10^{-21}\,s^{-1}$~\cite{Uzan}. We postpone a detailed account of the effect of the ELA metric in planetary orbits to another work, here we have considered that below the heliocentric distances of $20\,AU$
$\alpha(r_1<20\,AU)=3$~(\ref{alpha}) for which the physical gravitational acceleration is of order $\ddot{r}_p\sim 10^{-24}\,ms^{-2}$ such that $\dot{r}_{orb}/r_{orb}\sim -10^{-37}\,s^{-1}$, hence of the same order of magnitude of pure expansion effects.

As for the mass-energy densities~(\ref{rho}) we note that up to the heliocentric distance of $20\,AU$ it is
slightly below the cosmological quantity~$\rho_{RW}$, hence contributing a negative amount to the cosmological mass-energy density~(\ref{cosm_den}), $\rho_{(\alpha)}-\rho_{RW}<0$. Above the heliocentric distances of $70~AU$ the mass-energy densities for the fit $\alpha_{max}$ is exactly null due to saturating the bound~(\ref{max_bound}) while the remaining 3 fits, $\alpha_+$, $\alpha_-$ and $\alpha_{min}$ correspond to mass-energy densities well
above the cosmological quantity~$\rho_{(\alpha)}-\rho_{RW}\gg 0$, hence while $\alpha_{max}$ contributes a negative amount to the cosmological mass-energy density, $\alpha_+$, $\alpha_-$ and $\alpha_{min}$ contribute a positive amount.
Although the Pioneer experimental data only extends up to the heliocentric distance of $70\,AU$, for exemplification purposes of estimating maximal contributions to the relative cosmological mass-energy density $\Omega_\alpha$ we extend the fits obtained up to some heliocentric distance $R_{max}> 70\,AU$. Above $R_{max}$ we consider $\alpha=0$~(\ref{alpha}) for which the ELA metric coincides with the isotropic McVittie metric~\cite{McV1} such that spatial isotropy is exactly recovered and $\rho_{(\alpha)}(r_1>R_{max})=\rho_{RW}$, hence the contribution to $M_\alpha$ above $R_{max}$ due to the
ELA metric background is null. The contributions for the 4 fits of the functional parameter $\alpha$ represented in figure~\ref{fig.fits}, as well as the respective mass ratios $M_{\alpha}/M_{\odot}$, are plotted in figure~\ref{fig.rho} up to the heliocentric distance $100\,AU$.
\fig{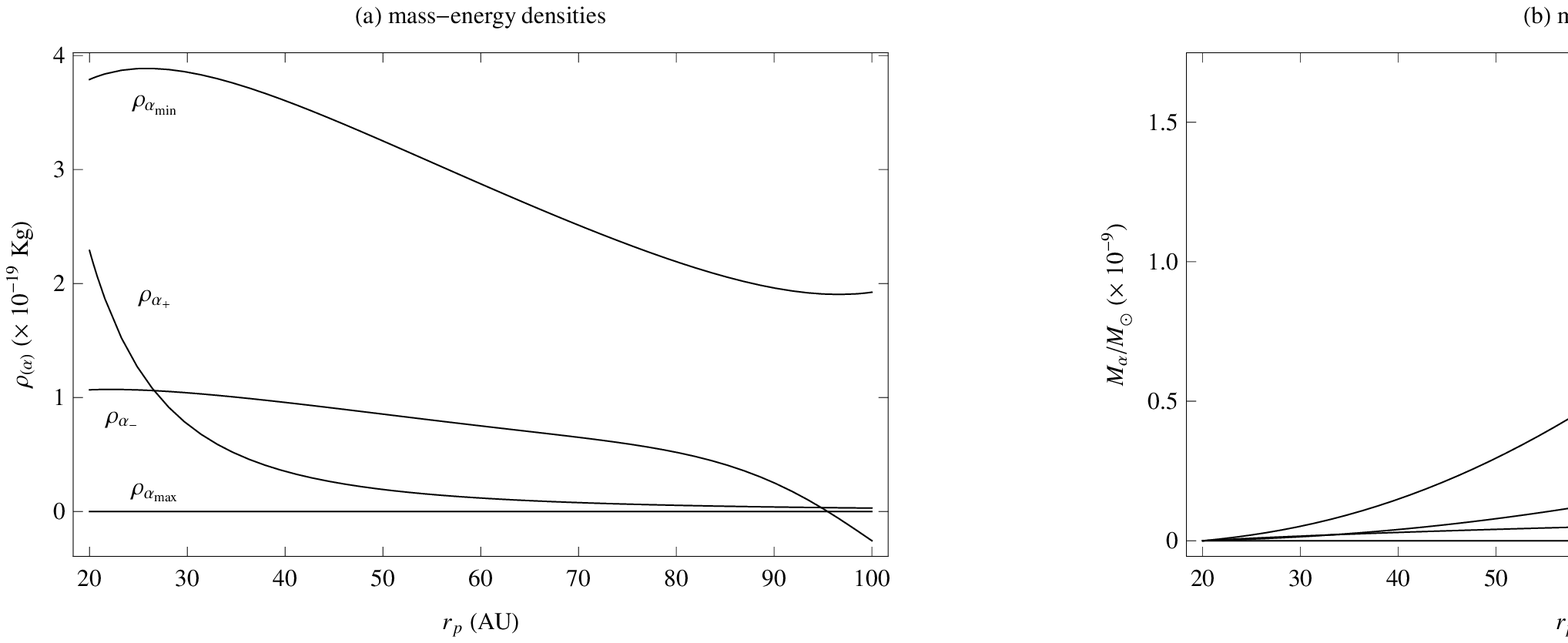}{90mm}{{\bf (a)} mass-energy densities $\rho_{(\alpha)}$ for the fits $\alpha_{max}$, $\alpha_+$, $\alpha_-$ and $\alpha_{min}$ represented in figure~\ref{fig.fits}; {\bf (b)} respective mass ratios $M_{\alpha}/M_\odot$, the negative mass ratio $M_{\alpha_{max}}/M_\odot\sim -10^{-17}$ is several orders of magnitude below the remaining fits not being represented in the figure.}{fig.rho}

Therefore, in the range of the heliocentric distances $r_p\in]20\,AU,70\,AU[$, the fits obtained
contribute at most an amount of order $\sim 10^{-9}$ to the relative cosmological mass-energy density $\Omega$. We note that, as already discussed in
the previous section, within the $\Lambda CDM$ model these contributions can only be included in the relative
cosmological mass-energy density of cold dark matter $\Omega_{CDM}=0.226$. Hence these contributions are negligible up to $70\,AU$.

For exemplification and analysis purposes, let us extend these fits above the Pioneer data range of $70\,AU$. For the fit $\alpha_{max}$ the mass-energy density is null, hence the contribution to $\Omega_\alpha$ is negative being given by $-\Omega_b\frac{4\pi\rho_{RW}}{3\,M_\odot}\,R_{max}^3$~(\ref{cosm_den}). This contribution can be fine-tuneable being unbounded from below, hence increasing the required value of contributions to $\Omega_{CDM}$ of other sources that not the ELA metric background. For the fit $\alpha_{-}$ the mass-energy density is positive above the heliocentric distance of $20\,AU$ up to spatial infinity decreasing monotonically with $r_p$ such that at $r_p=10930\,AU$ it exactly matches the cosmological quantity $\rho_{RW}$ and vanishes at spatial infinity. Hence the maximum positive relative contribution to the cosmological mass-energy density is $\Omega_\alpha=+6.412\times 10^{-10}\Omega_b$ corresponding to the cut-off $R_{max}=10930\,AU$. As for the maximum negative relative contribution to the cosmological density it is fine-tuneable and unbounded from below similarly to the fit $\alpha_{max}$. As for the fit $\alpha_{-}$, the mass-energy density $\rho_{\alpha_{-}}$ is positive above the heliocentric distance of $20\,AU$ up to approximately $r_1=95.48\,AU$ becoming negative above these distances such that the maximum allowed value of the ansatz cut-off~(\ref{alpha}) coincides approximately with this value, $R_{max}\lessapprox 95.48\,AU$. For this fit the contribution to the relative cosmological density is always positive having the maximum value of $\Omega_\alpha=+3.367\times 10^{-10}\Omega_b$ for $R_{max}=95.477\,AU$. Finally for the fit $\alpha_{min}$ the mass-energy density $\rho_{\alpha_{min}}$ is positive above $20\,AU$ increasing monotonically above the heliocentric distances of $96.596\,AU$. Hence the contribution to the relative cosmological mass-energy density is unbounded from above exactly matching the cold dark matter contribution $\Omega_\alpha=5\Omega_b=\Omega_{CDM}$ for $R_{max}=183.328\,AU$. These results are summarized
in table~\ref{table}.
\begin{table}[ht]
\begin{center}
\begin{tabular}[t]{llll}
Fit&$\rho$&$R_{max}$&$\Omega_\alpha$\\[2mm]\hline\\[-2mm]
$\alpha_{max}$&$=0$&fine-tunable&$=-\Omega_b\frac{4\pi\rho_{RW}}{3\,M_\odot}\,R_{max}^3<0$\\[4mm]
$\alpha_+$&$>\rho_{RW}$&$=10930\,AU$&$=+6.412\times 10^{-10}\Omega_b$\\
 &$(<\rho_{RW}$&$>10930\,AU$&$<+6.412\times 10^{-10}\Omega_b)$\\[4mm]
$\alpha_-$&$>\rho_{RW}$&$=95.477\,AU$&$=+3.367\times 10^{-10}\Omega_b$\\[4mm]
$\alpha_{min}$&$>\rho_{RW}$&$=183.328\,AU$&$=+5\Omega_b=\Omega_{CDM}$\\[2mm]\hline
\end{tabular}
\end{center}
\caption{Resume of maximal contributions to the relative cosmological mass-energy densities for the fits to $\alpha_{max}$, $\alpha_{+}$, $\alpha_{-}$ and $\alpha_{min}$.\label{table}}
\end{table}

From the above discussion we conclude that the fits $\alpha_{max}$ and $\alpha_{+}$ represented in figure~\ref{fig.fits}, corresponding to a positive Doppler shift in the range $r_p\in]20\,AU,70\,AU[$,
are compatible with the cosmological $\Lambda CDM$ model only when the fits are extended up to $\sim 10\,pc$.
When higher ranges for these fits are considered the negative contributions to the relative cosmological mass-energy densities become relevant such that the required relative (positive) density of cold dark matter to offset these contribution increases significantly hence becoming non-compatible with estimative for dark matter density in the universe~\cite{wmap,DM1}. As for the fits $\alpha_{-}$ and $\alpha_{min}$, corresponding to a negative Doppler
shift in the range $r_p\in]20\,AU,70\,AU[$ have more restrictive bounds being valid only up to order $\sim 10^2\,AU$. Above this range the mass-energy density corresponding to the fit $\alpha_{-}$ becomes strictly negative and for the fit $\alpha_{min}$ the relative mass-energy contribution exceeds the one
attributed to cold dark matter. Hence the fits $\alpha_{max}$ and $\alpha_{+}$ allow for a relatively
larger range than the fits $\alpha_{-}$ and $\alpha_{min}$.

For last let us discuss the equation of state corresponding to the functional parameter fits. We have
already concluded that the cosmological equation of state is not affected by the ansatz for $\alpha$~(\ref{alpha}), the average correction to the pressure is null~(\ref{ave_pressures}) and within
the $\Lambda CDM$ the corrections to the mass-energy may be included in the relative density of $\Omega_{CDM}$. However locally there are significant deviations from the cosmological equation
of state as already analysed in~\cite{PLB} for a constant parameter $\alpha$. Specifically for the
limiting case corresponding to the fit $\alpha_{max}$ the mass-energy density is exactly null from heliocentric distances of $20\,AU$ up to $R_{max}$, hence it is not possible to define a equation of state. As for the anisotropic equations of state $\omega_r$ and $\omega_\theta=\omega_\varphi$ for the fits $\alpha_{+}$, $\alpha_{-}$ and $\alpha_{min}$ are plotted in figure~\ref{fig.eq+}, figure~\ref{fig.eq-}
and figure~\ref{fig.eqmin}, respectively.
\fig{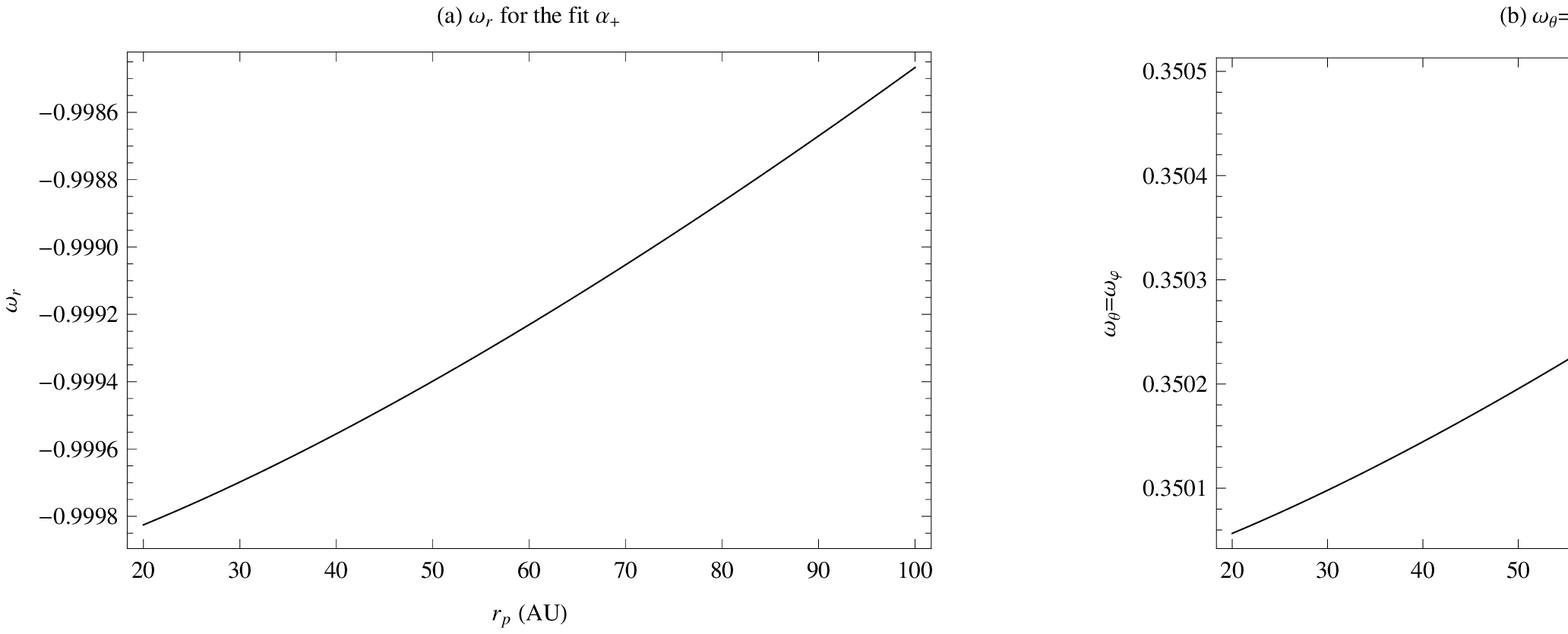}{90mm}{Equation of state for the fit $\alpha_+$ of the functional parameter $\alpha_2(r_1)$~(\ref{alpha}): (a) radial equation of state $\omega_{r}$; (b) angular equation of state $\omega_{\theta}=\omega_{\varphi}$.}{fig.eq+}
\fig{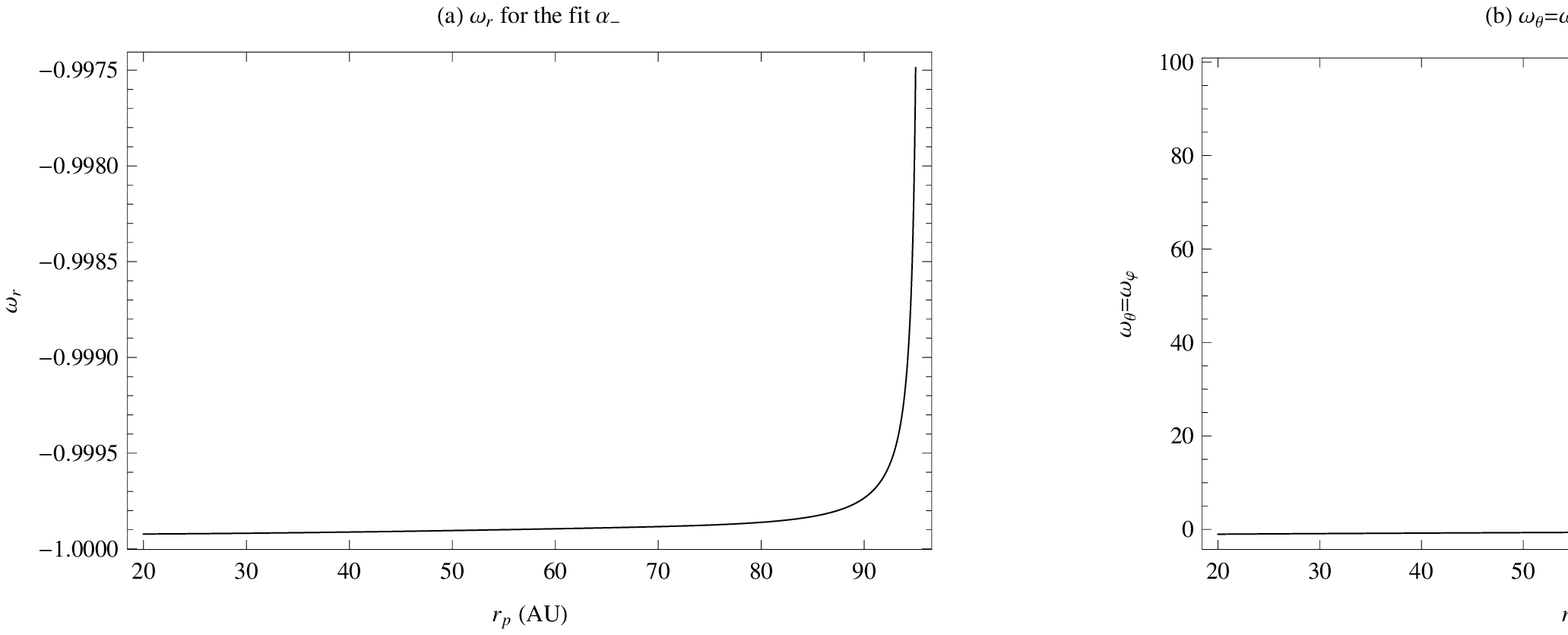}{90mm}{Equation of state for the fit $\alpha_-$ of the functional parameter $\alpha_2(r_1)$~(\ref{alpha}): (a) radial equation of state $\omega_{r}$; (b) angular equation of state $\omega_{\theta}=\omega_{\varphi}$.}{fig.eq-}
\fig{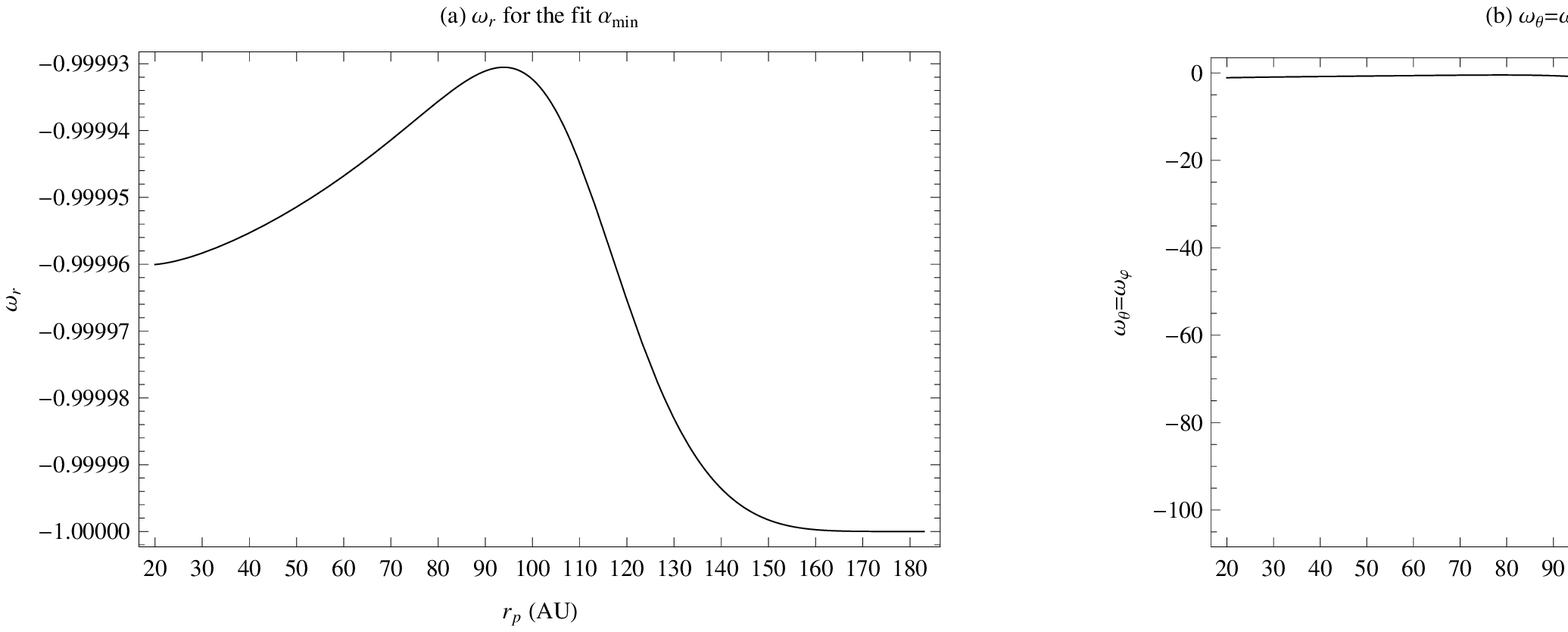}{90mm}{Equation of state for the fit $\alpha_{min}$ of the functional parameter $\alpha_2(r_1)$~(\ref{alpha}): (a) radial equation of state $\omega_{r}$; (b) angular equation of state $\omega_{\theta}=\omega_{\varphi}$.}{fig.eqmin}
For the fit $\alpha_{+}$ the radial equation of state $\omega_r$ above heliocentric distance of $20\,AU$ is above $-1$, being negative up to $12154.3\,AU$ and positive above this distance while the angular equation of state $\omega_\theta=\omega_\varphi$ is always positive. For the fit $\alpha_{-}$ the radial equation of state
$\omega_{r}$ is negative and above $-1$ in the range $r_p\in]20\,AU,R_{max}=95.48[$ while the angular equation of state is positive in this range. For the fit $\alpha_{min}$ the radial equation of state $\omega_r$ is negative and above $-1$ in the range $r_p\in]20\,AU,R_{max}=183.328\,AU[$ while the angular equation of state
is negative being below $-1$ in the range $r_p\in]20\,AU,25.86\,AU[\cup]96.60\,AU,R_{max}=183.328\,AU[$ and
above $-1$ in the range $r_p\in]25.86\,AU,96.60\,AU[$.

As for the possible interpretation for the matter content corresponding to the several fits discussed we note that the fits $\alpha_{+}$ and $\alpha_{-}$ have equation of state compatible with a standard gauge field ($\omega_{r(\alpha)}>-1$ and $\omega_{\theta(\alpha)}=\omega_{\varphi(\alpha)}>-1$) while the fit $\alpha_{min}$ has a equation of state along the radial
direction compatible with a standard gauge field ($\omega_{r(\alpha)}>-1$) as well as along the angular variable directions in the range $r_p\in]25.86\,AU,96.60\,AU[$ while in the range $r_p\in]20\,AU,25.86\,AU[\cup]96.60\,AU,R_{max}=183.328\,AU[$ it has a equation of state that resembles Phantom matter ($\omega_{\theta(\alpha)}=\omega_{\varphi(\alpha)}<-1$).

\section{Conclusions}

In this work we have computed the measurable Doppler acceleration corrections
for a background described by the ELA metric~(\ref{ELA_metric}) and fitted
an ansatz for the functional parameter of this metric~(\ref{alpha})
to the current Pioneer anomaly bounds on the range $r_p\in]20\,AU,70\,AU[$, hence
setting lower bounds for the functional parameter $\alpha$ in this range.

In particular we have consider the 4 fits represented in figure~\ref{fig.fits} within the error bars for
the Pioneer Doppler acceleration, the fit $\alpha_{max}$ corresponding to the limiting solution~(\ref{max_bound}),
the fit $\alpha_{+}$ corresponding to $9/10$ of the limiting solution~(\ref{max_bound})
and the power series~(\ref{polinomial_fit}) fits $\alpha_{-}$ and $\alpha_{min}$.
The main contribution to the Doppler acceleration corresponding to these fits is due to the background gravitational frequency-shift~(\ref{F_Doppler}),
being of order $\delta\ddot{r}_p^{Doppler}\sim 10^{-10}\,ms^{-2}$ while the physical gravitational acceleration~(\ref{F}) is lower by 4 orders
of magnitude, $\delta\ddot{r}_p\sim 10^{-14}\,ms^{-2}$.

We have also computed the mass-energy density and anisotropic pressures for the ELA metric~(\ref{rho})
and analysed the possible contributions due to this background to the relative cosmological mass-energy
densities and pressures. There are no contributions for the cosmological pressure, however extrapolating the fits above the heliocentric distance of $70\,AU$ are obtained significant contributions to the cosmological
mass-energy density. For compatibility with the $\Lambda CDM$ model we conclude that these corrections
must be included in the mass-energy density attributed to cold dark matter $\Omega_{CDM}$ such that the
cosmological equation of state is not modified. Hence we note that the ELA metric background may constitute
a covariant parameterization for cold dark matter. We also discuss the anisotropic local equation of state
for the fits obtained noting that, although along the radial coordinate direction the equation of state is above the value $-1$, hence being
compatible with the equation of state for a standard scalar field, along the angular variables directions, depending on the heliocentric distance $r_p$, the equation of state is either above, either below the value $-1$, hence being either compatible with the equation of state for a standard scalar field, either compatible with the equation of state for a Phantom field.

In addition let us recall that the original motivation for considering the existence of either cold dark matter, either gravitational interaction corrections is the observed flattening of galaxies rotation curves~\cite{DM1} which is equivalent to a significant correction to the gravitational acceleration towards the centre of the galaxies growing with the radial distance to the centre core of the galaxies. With respect to our fits we note that, for heliocentric distances above $70\,AU$ up to $\sim 10\,pc$ the gravitational acceleration correction for the fits $\alpha_{max}$ and $\alpha_{+}$ is exclusively towards the Sun decreasing monotonically in absolute value. As for the fit $\alpha_{-}$ the gravitational acceleration correction is exclusively outwards the Sun up to the heliocentric distance $R_{max}=95.477\,AU$ and for the fit $\alpha_{min}$ it is outwards the Sun up to $91\,AU$ turning towards the Sun above this distance having a pronounced peak at $113\,AU$.
Hence, if the ELA metric is interpreted as a covariant parameterization of cold dark matter, a extrapolation to galaxy scales with higher values for the central mass, tends to single out the fit $\alpha_{min}$ as the only fit that simultaneously allows for a positive contribution to $\Omega_{CDM}$ and a gravitational acceleration correction towards the central mass at large radial distances, excluding the fit $\alpha_{-}$ and still allowing both the fits $\alpha_{max}$ and $\alpha_{+}$ although not justifying the contribution for $\Omega_{CDM}$. We will develop such study in another work~\cite{curves}.

\section*{Acknowledgments}
Work supported by grant SFRH/BPD/34566/2007 from FCT-MCTES. Work developed in the scope of
the strategical project of GFM-UL PEst-OE/MAT/UI0208/2011.

\appendix

\section{Appendix\label{sec.app}}

The coefficient values for the power series fit~(\ref{polinomial_fit}) obtained by a least square
minimization method to $\alpha_{-}$ are
\begin{equation}
\begin{array}{rclcrcl}
\alpha_{2.-1}&=&+1.31715\times 10^{20}\,m&,& \alpha_{2.0}&=&+5.37413\times 10^8\ ,\\
\alpha_{2.1}&=&-0.00612127\,m^{-1}&,&\alpha_{2.2}&=&+1.07833\times 10^{-17}\,m^{-2}\ ,\\
\alpha_{2.3}&=&+1.5718\times 10^{-30}\,m^{-3}&,&\alpha_{2.4}&=&-1.07371\times 10^{-43}\,m^{-4}\ ,\\
\alpha_{2.5}&=&+3.98678\times 10^{-57}\,m^{-5}&,&\alpha_{2.6}&=&+1.59514\times 10^{-70}\,m^{-6}\ ,\\
\alpha_{2.7}&=&-4.3892\times 10^{-83}\,m^{-7}&,&\alpha_{2.8}&=&+3.39347\times 10^{-96}\,m^{-8}\ ,\\
\alpha_{2.9}&=&-9.79789\times 10^{-110}\,m^{-9}
\end{array}
\end{equation}
and to $\alpha_{min}$ are
\begin{equation}
\begin{array}{rclcrcl}
\alpha_{2.-1}&=& - 1.33106\times 10^{20}\,m&,& \alpha_{2.0}&=&5.69474\times 10^8\ ,\\
\alpha_{2.1}&=&- 0.00573066\,m^{-1}&,&\alpha_{2.2}&=&+ 2.66023\times 10^{-17}\,m^{-2}\ ,\\
\alpha_{2.3}&=& 3.59912\times 10^{-31}\,m^{-3}&,&\alpha_{2.4}&=&- 1.69481\times 10^{-43}\,m^{-4}\ ,\\
\alpha_{2.5}&=&+ 1.18619\times 10^{-56}\,m^{-5}&,&\alpha_{2.6}&=& 9.43547\times 10^{-70}\,m^{-6}\ ,\\
\alpha_{2.7}&=&- 7.82415\times 10^{-83}\,m^{-7}&,&\alpha_{2.8}&=&- 6.14959\times 10^{-96}\,m^{-8}\ ,\\
\alpha_{2.9}&=& 4.64438\times 10^{-109}\,m^{-9}
\end{array}
\end{equation}

\label{lastpage}

\end{document}